\documentclass[12pt,preprint]{aastex}

\shorttitle{Activity index distribution}
\shortauthors{Zhao et al.}

\begin{document}


\title{TRACERS OF CHROMOSPHERIC STRUCTURE. I. CaII H$\&$K EMISSION DISTRIBUTION OF 13000 F, G AND K STARS IN SDSS DR7 SPECTROSCOPIC SAMPLE}
\author{J. K. Zhao\altaffilmark{1,2}, T. D. Oswalt\altaffilmark{2}, G. Zhao\altaffilmark{1}, Q. H. Lu\altaffilmark{3}, A. L. Luo\altaffilmark{1}, L. Y. Zhang\altaffilmark{4}}


\altaffiltext{1}{Key Laboratory of Optical Astronomy, National Astronomical Observatories, Chinese Academy of Sciences, Beijing, 100012, China; zjk@bao.ac.cn, gzhao@bao.ac.cn, lal@bao.ac.cn}
\altaffiltext{2}{Physics and Space Science Department, Florida Institute of Technology, Melbourne, USA, 32901; toswalt@fit.edu}
\altaffiltext{3}{Department of Information Engineering Jiangsu Animal Husbandry $\&$ Veterinary College, No. 8 Fenghuang Road, Taizhou,Jiangsu Province, China; tzlqh2006@163.com}
\altaffiltext{4}{Guizhou University, Guiyang, Guizhou, P.R.China; liy\_zhang@hotmail.com}


\begin{abstract}
We present chromospheric activity index $S\rm_{HK}$ measurements for over 13,000 F, G and K disk stars with high signal-to-noise ratio ($>$ 60) spectra in the Sloan Digital Sky Survey (SDSS) Data Release 7 (DR7) spectroscopic sample.  A parameter $\delta$S is defined as the difference between $S\rm_{HK}$ and a `zero' emission line fitted by several of the most inactive stars. The $S\rm_{HK}$ indices of subgiant stars tend to be much lower than dwarfs, which provide a way to distinguish dwarfs and giants with relatively low resolution spectra. Cooler stars are generally more active and display a larger scatter than hotter stars. Stars associated with the thick disk are in general less active than those of the thin disk. The fraction of K dwarfs that are active drops with vertical distance from the Galactic plane. 
  Metallicity affects $S\rm_{HK}$ measurements differently among F, G and K dwarfs in this sample. Using the open clusters NGC 2420, M67 and NGC6791 as calibrations, ages of most field stars in this SDSS sample range from 3-8 Gyr.

\end{abstract}



\section{Introduction}
 Among solar-type stars, chromospheric activity (CA) is closely related to the
stellar dynamo, magnetism and rotation rate (Middelkoop
1982a/b; Rutten 1984).  It is generally accepted that
magnetic activity in late-type stars is the product of an $\alpha$-$\Omega$ dynamo,
which results from the action of differential rotation at the
tachocline (the interface between the convective envelope and
the radiative core).

The most common indicator of CA
is the well-known $S$ index, defined as the ratio of the flux
in the core of the Ca II H$\&$K lines to the nearby continuum
(Vaughan et al. 1978). Early work by Wilson (1963; 1968) and Vaughan $\&$ Preston (1980) established CaII H$\&$K emission as a useful marker of CA in lower main sequence (MS) stars. In late F to early M dwarfs Skumanich (1972) found that CaII H$\&$K emission, magnetic field strength and rotation all decay as the inverse square root of stellar age. Thus, CA is a potential age indicator and several efforts have been undertaken to calibrate
it (see e.g. Soderblom et al. 1991; Lachaume et al. 1999;
Mamajek $\&$ Hillenbrand 2008, Soderblom 2010 and references therein).

 Stellar CA has been studied in large samples of nearby ($<$ 200pc) stars
(Duncan et al. 1991;
Henry et al. 1996; Gray et al. 2003, 2006; Wright 2004; Jenkins et al. 2006, 2008, 2011; Arriagada 2011). The Sloan Digital Sky Survey (SDSS; York et al. 2000) has obtained more than 464,000 stellar spectra. Most of these stars are more distant than 200 pc. West et al. (2004; 2008) investigated CA among M dwarfs using SDSS spectra. They showed the fraction of magnetically active stars (as traced by H$\alpha$ emission) decreases as a function of the vertical distance from the Galactic plane. Considering that active stars are generally younger than inactive stars, West et al. interpreted this in terms of thin-disk dynamical heating and a rapid decrease in magnetic activity.

Zhao et al. (2012) measured Ca II H$\&$K emission lines of 80 wide binary candidates using SDSS spectra and found that CA levels of components in a binary are similar. This supports the assumption that such wide binaries are coeval.

Rocha-Pinto $\&$ Maciel (1998) found a relationship between the `age excess', defined as the difference between
the stellar isochrone and chromospheric ages, and the metallicity, as measured by the index
[Fe/H] among late-type dwarfs. The chromospheric age tends to be younger than the isochrone age
for metal-poor stars; the opposite occurs for metal-rich stars. Gray et al. (2006) pointed out that metallicity can affect the CA distribution and therefore the $S$ index.  A tail of quite active stars persists in the CA distribution even at quite low
metallicities ([Fe/H] $<$ -0.5). They also suggested the Noyes et al. (1984) calculation of log $R\arcmin\rm_{HK}$ should include a correction factor for this metallicity effect.


In this paper, we describe a study of the CA distribution among a large sample of F, G and K stars. Our main goal is to investigate the relation between average activity, activity fraction and vertical distance from the Galactic plane. In addition, we examined the metallicity effect on the activity index. Using our results, we roughly estimated the age of the total spectroscopic sample of SDSS stars.

It is well known that the binary fraction among F and G starts is quite high. Some unresolved binaries show strong emission of chromospheric activity. Our study can not address the issue of binarity. So we must accept this will be a major source of scatter in our CA vs. age relations.

Section 2 presents a discussion of our data selection criteria. The measurements of $S\rm_{HK}$ and analysis of the CA distribution are presented in Section 3. Section 4 gives a discussion of our results. Our conclusions are presented in Section 5.
\section{DATA}


\subsection{Overview of the SDSS Spectroscopic Data}
The SDSS provides a homogeneous and deep ($r$ $<$ 22.5) photometric sample in five band passes ($u$, $g$, $r$, $i$, and $z$; Gunn
et al. 1998, 2006; Hogg et al. 2001; Smith et al. 2002; Tucker
et al. 2006). It is accurate to $\pm$0.02 mag (rms scatter) for unresolved
sources not limited by photon statistics (Scranton et al. 2002) and a zero-point uncertainty of $\pm$0.02 mag (Ivezi\'{c} et al. 2004). The SDSS also provides more than half a million stellar spectra spanning 3800 - 9000 $\rm \AA$. Radial Velocity (RV), metallicity [Fe/H], $T$$\rm_{eff}$ and log $g$ are provided in the Table \textit{sppParams} (Lee et al. 2008). SDSS spectroscopy was carried out using twin fiber-fed spectrographs collecting 640 simultaneous observations. Typical exposure times were $\sim$15 - 20 minutes, but exposures were subsequently co-added for total exposure times of
$\sim$45 minutes, producing low resolution spectra with R $\sim$
2000 (York et al. 2000). SDSS spectroscopic plates each contained
16 spectrophotometric standard stars, which were selected
by color to be subdwarf F stars. Absolute fluxes were calibrated by comparing these standard stars to a grid
of theoretical spectra from  model atmospheres (Kurucz
1993) and obtaining a spectrophotometric solution for
each plate.

\subsection{ Data Selection}
We selected F, G and K stars with high S/N from the SDSS archive. The sample was initially extracted mainly based on color: 0.3 $<$ $(g-r)_{0}\footnote[1]{The subscript nomenclature means dereddened color.}$ $<$ 1.3.  To eliminate M stars, we added other color constraints, i.e.,  $i-z$ $<$ 0.3 and $r-i$ $<$ 0.53. Lower metallicity stars were excluded by the constraint: [Fe/H] $>$ -1.0. For stars older than 1 Gyr, the Ca II emission lines are very weak, so high S/N  spectra are needed. Therefore, only stars with S/N $>$ 60 were included in our sample. With the above constraints, 13,198 stars were culled. In order to estimate photometric parallaxes of dwarf stars, we adopted the relation from Ivezi\'{c} et al. (2008), which gives the absolute magnitude in the $r$ band, $M\rm_{r}$ as a function of \textit{g-i} and \textit{[Fe/H]}. Then, using the positions and distances of each star, the Galactic height was computed assuming the Sun is 15 pc above the plane. We used the SDSS/USNO-B-matched catalog (Munn et al. 2004) to obtain proper motions.

 Fig. 1 presents the reduced proper motion (RPM) diagram for this sample, which offers a powerful tool for segregating members of kinematically-distinct stellar populations. The x-axis is $(g-i)_{0}$. The y-axis is $H\rm_{r}$ $\equiv$ $r$+5*log$\mu$+5, where $\mu$ is the proper motion in arcseconds/year. The dashed line indicates the division between disk and halo stars from Smith et al. (2009). Asterisks represent possible halo stars. It is clear that nearly all the stars in our sample are members of the disk. The gaps and overdense regions in Fig. 1 do not represent the actual distribution of stars but are a reflection of the SDSS spectroscopic targeting selection criteria. Fig. 2 is the metallicity distribution of our sample. The metallicity [Fe/H] can be estimated from SDSS spectra using several methods (Lee et al. 2008). The peak of our sample is at [Fe/H] $\sim$ -0.4. In general, the average metallicity of our sample is lower than that of Nordstr\"{o}m et al. (2004). We would expect a more metal poor population in our sample because the vertical distance from the Galactic plane of our sample is larger than that of Nordstr\"{o}m et al. (2004) whose mean vertical distance is smaller than 100 pc. 

\subsection{Open Clusters in the SDSS}
To estimate the ages of stars in our sample, three open clusters NGC2420, M67 and NGC6791 were selected from the SDSS DR7 as baselines for an $S$$\rm_{HK}$ vs. age relation. The selection method for cluster member stars was almost the same as that in Zhao et al. (2012) except that the latter selected cluster stars from the DR8, while in this paper the DR7 was used. The age of M67 is about 4.05 Gyr (Jorgensen $\&$ Lindegren 2005). The age of NGC6791 is about 8 Gyr (Grundahl et al. 2008). The age of NGC2420 is about 2.0 Gyr (Von Hippel $\&$ Gilmore 2000).

\section{Analysis}
\subsection{$S\rm_{HK}$ Measurement }
 Before spectral index measurements were made our spectra were corrected for radial velocities. Then, following Hall et al. (2007), for each star we computed the flux ratio $S$$\rm_{HK}$:
\begin{eqnarray}
S\rm_{HK}&\equiv&\rm{\alpha\frac{H+K}{R+V}}
\end{eqnarray}
where H and K are the fluxes measured in 2 $\rm{\AA}$ rectangular windows centered on the line cores of CaII H (centered at 3968.5$\rm \AA$) $\&$ K (centered at 3933.7$\rm \AA$); R and V  centered at 3901$\rm \AA$ and 4001$\rm \AA$ are the fluxes measured in 20 $\rm{\AA}$ rectangular `pseudocontinuum' windows on either side. Detailed discussion of $S\rm_{HK}$ measurements can be found in Zhao et al. (2011).  Fig. 3 shows the spectra of an active star (dotted line) and an inactive star (solid line). Short solid lines show the wavelength range of the H, K, R and V bandpasses. The emission reversals sensitive to CA are indicated by arrows. Clearly, the emission line of CaII H$\&$K can be used to investigate the stellar activity using SDSS spectra.

To account for different continuum flux levels near the CaII lines for stars of different spectral type, the $S\rm_{HK}$ values are often parameterized as log $R\arcmin\rm_{HK}$, the ratio of the flux in the H and K line cores to the total bolometric emission of the star. As in previous work (Henry et al. 1996; Gray et al. 2003; Wright et al. 2004; Jenkins et al. 2011), stars were observed in common with the Mount Wilson sample. $S$ values were then transformed to the equivalent Mount Wilson $S$ values. Then, following the formula of Noyes et al. (1984), log $R\arcmin\rm_{HK}$ values were calculated. A disadvantage of using log $R\arcmin\rm_{HK}$ is that stars with B-V $\leq$ 0.4 or $\geq$ 1.0 were not included in the original calibrations because they were not observable in distant clusters (Noyes et al. 1984; Mamajek $\&$ Hillenbrand
2008). Thus, the derived quantities log$R\arcmin\rm_{HK}$, P$\rm_{rot}$, and stellar age are most secure for MS stars that have log $R\arcmin\rm_{HK}$  between -4.0 and -5.1 and B-V values between 0.4 and 1.0. Since the stars in our sample are very faint, none are in the Mount Wilson sample. Thus, to remove the effect of the photospheric contribution, we chose to use an instrumental $\delta$S defined below (see Fig. 4). It should be understood that for these least active stars they still may be weak emission that can not be detected at the resolution afforded by SDSS spectra.

\subsection{CA distribution}
Vaughan $\&$ Preston (1980) measured H $\&$ K lines for 486 F-G-K-M MS stars and found  that the relative numbers of more-active (Hyades-like) and less-active (solar-like) F-G stars are roughly consistent with a nearly constant rate of star formation. However, there exists an apparent deficiency in the number of F-G stars exhibiting intermediate activity, which has been dubbed the `Vaughan-Preston gap'. Henry et al. (1996) presented CA measurements of 800 southern stars within 50 pc. They also found a bimodal distribution of stellar CA in their sample. Gray et al. (2003; 2006) presented the CA of 3600 dwarf and giant stars earlier than M$\rm{0}$ within 40 pc of the sun. They demonstrated that the chromospheric emission
parameter log $R\arcmin\rm_{HK}$ has a bimodal distribution, which were interpreted by them as a manifestation of the Vaughan-Preston gap. However, they suggested that this bimodality is dependent on metallicity. For stars with [Fe/H] $>$ -0.20, the distribution is bimodal, but the distribution is strictly single-peaked for stars of lower metallicity. Isaacson $\&$ Fischer (2010) presented measurements of CA for more than 2600 MS and subgiant stars in the California Planet Search (CPS) program with spectral types ranging from about F5V to M4V for
MS stars and from G0IV to about K5IV for subgiants. They showed the vast majority of subgiants have very low chromospheric activity.

\subsubsection{CA distribution of total sample}
Fig. 4 presents the $S\rm_{HK}$ vs. $(g-r)_{0}$ relation for our total SDSS sample. As above, the `0' subscript means dereddened color. Each black point represents a star.  The $S\rm_{HK}$ ranges from 0.1 to 0.7.  The solid  line is the CA `zero' emission line: it is a fourth order polynomial fit using ten filled circle points chosen by eye which are the least active stars in each color bin.  This `zero' emission line can be roughly regarded as the effect of photospheric contamination light on $S\rm_{HK}$, similar to that defined by Isaacson $\&$ Fischer (2010). Their `zero' emission line is a function of B-V and $S\rm_{HK}$. It has the similar shape to our `zero' emission line.  $\delta$S, defined as the difference between star's $S\rm_{HK}$ and the `zero' emission line, was adopted as our relative instrumental index of CA.

Our sample includes giants and dwarfs. The line profiles of giants are narrower than those of dwarfs. Generally $S\rm_{HK}$ among giants is known to be relatively lower than those of dwarfs. We choose a lower threshold of $S\rm_{HK}$ $\leq$ 0.17 to define the region populated by giants (dashed line in Fig. 4). Fig. 5 presents the log $g$ distribution for the two groups above and below this threshold.  The bottom panel is the log $g$ distribution of stars with $S\rm_{HK}$ $<$ 0.17. Log $g$ was found to be less than 4.0 for about 91\% of these stars, implying they are subgiants or giants. This is consistent with Isaacson $\&$ Fischer (2010), who also suggested that most subgiants have low CA. In their sample, only 10\% of subgiants were active and some of the active subgiants were found to be rapid rotators or close binaries. The top panel of Fig. 5 is our distribution for more active stars with $S\rm_{HK}$ $>$ 0.17.  For 88\% of these stars log $g$ $>$ 4.0 implying they are dwarf stars. Since log $g$ is a more difficult parameter to determine from low resolution spectra of faint stars than other parameters such as $T$$\rm_{eff}$ and [Fe/H], $S\rm_{HK}$ may be used to distinguish dwarfs and giants.

Fig. 6 presents the relation between mean $\delta$S and $(g-r)_{0}$ using only dwarf stars (log $g$ $<$ 4.0). The triangles indicate the mean $\delta$S of each color bin (binsize = 0.1 in $(g-r)_{0}$) and the error bars are the corresponding standard deviations. Fig. 6 shows that the mean $\delta$S as well as the scatter in $\delta$S among red stars are both larger than among blue stars. In other words, red stars have a broader range in CA, which supports that later type stars have longer active lifetimes.

Fig. 7 shows the $\delta$S distribution among stars as a function of color index. Among stars with 0.3 $<$ $(g-r)_{0}$ $<$ 0.9, only a small fraction of stars are active and there is only one peak in the distribution. The histograms for stars with $(g-r)_{0}$ $>$ 1.0 are clearly bimodal. The stars in this latter color range tend to be K type dwarfs with a broad range in CA.

\subsubsection{Activity difference in thin disk and thick disk }

To investigate the activity in different population, we compared the mean CA level in each color bin for thin disk stars and thick disk stars. Thin disk
stars and thick disk stars are mainly distinguished by their kinematics. The U, V and W space velocities of each star are calculated following the procedure in Zhao et al. (2011). Then,
we used Equations 1-3 of Bensby et al. (2003) and the
characteristic velocity dispersions and asymmetric drift values
given in their Table 1 to determine the likelihood for each star
of belonging to either the thin disk, thick disk based on
its kinematics alone.  Combining the kinematic and local number density
likelihoods, for each individual star we determine the likelihood
of it belonging to the thin disk, thick disk. Ratios of
these likelihoods were then used to find stars that are more
likely to be thin disk than thick disk (i.e., TD/D $<$ 0.10) and
more likely to be thick disk than thin disk (i.e., TD/D $>$
10).

Fig. 8 demonstrate the mean CA level in each color bin for thin disk stars and thick stars. In Fig. 8, the triangles indicate $<$$\delta$S$>$ for thin disk stars. Filled circles represent $<$$\delta$S$>$ for thick disk stars. The mean standard deviation for each are indicated by error bars. As seen in Fig. 8, the CA among thin disk stars tends to be stronger than among thick disk stars. For stars with  $(g-r)_{0}$ $<$ 0.6 (F and G stars), the mean difference in $<$$\delta$S$>$ between thin disk and thick disk is small, while the difference is more evident for stars with $(g-r)_{0}$ $>$ 0.6 (K dwarfs). 

\subsubsection{Activity and  vertical distance from Galactic plane}


West et al. (2008) investigated the relation between fraction of active stars and vertical distance of Galactic plane using the M dwarfs. We defined an active K dwarfs: $\delta$S $>$ 0.15.  To reduce the metallicity effect, two K  dwarf groups were selected based on their [Fe/H]. A high metallicity group was defined by [Fe/H] $>$ -0.5; a low metallicity group was defined by [Fe/H] $<$ -0.5. Fig. 9 shows the fraction of active K dwarfs vs. vertical distance {for high metallicity group, low metallicity group and all stars}. This fraction of active stars clearly decreases with vertical distance in each of these groups. More active stars are seen to be located close to Galactic plane. These stars are relative young. Because of thin-disk dynamical heating, the older stars are more likely to be found farther from the Galactic plane.

\subsection{CA and [Fe/H]}

Fig. 10 shows the relation between $S\rm_{HK}$, $(g-r)_{0}$ and [Fe/H] among F, G and K {dwarfs}. The bin size of each rectangle is 0.04 in  $(g-r)_{0}$ and 0.01 in $S\rm_{HK}$. The color bar to the right of Fig. 10 gives the relation between [Fe/H] and color. There is no evident correlation between $S\rm_{HK}$ and [Fe/H]. Among F dwarfs, there is a trend suggesting lower metallicity stars have stronger CA. It can also be seen that most active G dwarfs are metal rich stars. However, no trend can be seen among K dwarfs. In our sample, a clear relation between $S\rm_{HK}$ and [Fe/H] can not be derived. We only found the relation between $S\rm_{HK}$ and [Fe/H] for F $\&$ G stars is different from K dwarfs.

%

\subsection{Age of the sample}
Stellar age is one of the most difficult properties of a star to
determine. 
Pace et al. (2009) believed stars change from active to inactive,
crossing so-called Vaughan Preston gap on a timescale as short
as 200 Myr. 
Jenkins et al. (2011) and Zhao et al. (2011) showed that the age-activity relationship appears to extend to ages older than the Sun. Our sample in this paper provides
a chance to investigate the relation between age and CA.

We have identified member stars of three open clusters based on the cluster position, color magnitude diagram (CMD), radial velocity and [Fe/H] using the criteria from Smolinski et al. (2011). In Fig. 11, the open squares, blue filled circles and open triangles represent the member stars of the open clusters NGC2420, M67 and NGC6791, respectively. The dotted line, the dashed line and the dash-dot line are the respective least square fits and $\pm$1 $\sigma$ of those three clusters. The scatter in NGC2420 is large because the S/N of these spectra is low. The NGC6791 fitted line is concave down because this evolved cluster has no blue stars on the upper main sequence to define its curvature. Nevertheless, among these three open clusters it is still clear that CA decays with age to at least 8 Gyr. Our results contradict the suggestion by Pace et al. (2009) that CA remains constant after $\sim$ 1 Gyr.

The fitted lines of these three clusters can be used to roughly estimate the age of the field stars in our sample. For stars with 0.3 $<$ (g-r)$_{0}$ $<$ 0.47, ages range from 2 - 4 Gyr; for stars with 0.47 $<$ (g-r)$_{0}$ $<$ 0.6, ages range 2-8 Gyr; for stars with 0.6 $<$ (g-r)$_{0}$ $<$ 1.2, the ages are all younger than 8 Gyr. It should be noted that this age calibration is applicable only to the SDSS sample  because $S\rm_{HK}$ is defined here by the SDSS instrumental system.

\section{Discussion}
 Stellar activity is a collective term used to describe phenomena related to the presence of magnetic fields in cool stars, i.e. stars of spectral types ranging from late A to late M at the low-mass end of the main-sequence. Different activity distributions are seen among different spectral type stars. Many late type stars have strong surface magnetic fields (Johns-Krull $\&$ Valenti 1996) that heat the outer atmosphere above the photosphere and lead to observable emission from the chromosphere (e.g., Ca II and H Balmer series lines).  As previously shown in Fig. 6, late type stars also have relatively stronger average activity in our sample. The reason is in the stellar interior late type stars have larger convection zone. The CA distributions of late type stars are much more complicated than early type stars. As shown in Fig. 7,  within 0.3 $<$ $(g-r)_{0}$ $<$ 1.0 the distributions show only one peak, while within 1.0 $<$ $(g-r)_{0}$ $<$ 1.2 the distributions are bimodal. The physics that controls the production of magnetic fields in late type stars is still not well understood. The lack of a radiative-convective boundary layer seems to preclude storing large-scale fields as in a solar-type dynamo.

It is well known that there is a relation between CA and age. Generally, older stars have relatively weaker activity. Thus, activity can be used to trace galactic evolution.
Fig. 8 compared the CA distribution in the thin disk and thick disk. Stars in the former population have stronger activity because of their younger ages. Stars in the thin disk or thick disk are also distinguished by their kinematics. In view of this, CA also provides a proxy for kinematics. Fig. 9 presents the relation between the fraction of active stars and vertical distance from the Galactic plane among K dwarfs. Clearly, the fraction of active stars decreases with vertical distance. This is consistent with the result of  West et al. (2008) who found among 38,000 low mass dM stars that the fraction of magnetically active stars (as traced by H¦Á emission) decreases with the vertical distance from the Galactic plane.

Metallicity also appears to play a role in stellar activity. Rocha-Pinto $\&$ Maciel (1998) found a relationship between the `age excess', defined as the difference between
the stellar isochrone and chromospheric ages, and the metallicity, as measured by the index [Fe/H] among late-type dwarfs. However, in our sample, no clear CA and metallicity relation was found. Only a weak correlation is seen for stars within 0.3 $<$ (g-r)$_{0}$ $<$ 0.4. The probable reason is that the metallicity of our sample is relatively low compared to the sample of Rocha-Pinto $\&$ Maciel (1998).

Our results contrast those of Pace (2013) who found that CA drops abruptly at age $\sim$1.5 Gyr. Our SDSS sample clear shows a much more gradual decline in CA with age that extends to at least $\sim$8 Gyr. Pace (2013) does acknowledge that CA values among MS pairs in even old wide visual binaries are consistent. The latter could only be true if CA is a continuous fraction of age and spectral type (mass).

\section{Conclusion}
We selected a sample of over 13,000 F, G and K stars from the SDSS DR7 spectroscopic archive. As expected, this sample was found to be nearly disk stars. The CA index $\delta$S was measured for these stars, calculated as the difference between $S\rm_{HK}$ and  a `zero' emission line that defines the least active stars envelope across all $(g-r)_{0}$ color. The measured CA range among red stars was found to be much wider than among blue stars. We also found that the fraction of active K dwarfs decreases with vertical distance from the Galactic plane. However, among F $\&$ G stars, there is no obvious difference in activity between the thin disk stars and thick disk stars. Thin disk K dwarfs tend to have much stronger CA than thick disk K dwarfs.

With the help of three open clusters in an SDSS sample, we roughly estimated relative ages. For stars with 0.3 $<$ $(g-r)_{0}$ $<$ 0.47, typical ages ranged from 2 - 4 Gyr, while ages ranged from 2-8 Gyr for stars with 0.47 $<$ $(g-r)_{0}$ $<$ 1.2.

Metallicty is known to affect CA differently among F, G and K dwarfs. We found no evident relation between CA and metallicity among G and K dwarfs, yet some F dwarfs seem to have stronger CA only because of their lower metallicity or stronger photospheric contamination.

The CA values of subgiants and giants tend to be much smaller than those of dwarfs in accord with earlier studies. Thus, such measurements may be used to distinguish dwarfs and giants even at the low resolution of the SDSS spectra.

%




\acknowledgments
We are grateful for constructive comments by the reviewer that substantially improved our paper.
TDO acknowledges support from NSF grant AST-0807919 to Florida Institute of Technology. JKZ and GZ acknowledges  support from NSFC grant No. 11078019 and 11233004.  LYZ acknowledges  support from NSFC grant No. 10978010. LAL acknowledges  support from NSFC grant No. 10973021.

Funding for the SDSS and SDSS-II has been provided
by the Alfred P. Sloan Foundation, the Participating Institutions,
the National Science Foundation, the US Department of
Energy, the National Aeronautics and Space Administration,
the Japanese Monbukagakusho, the Max Planck Society, and
the Higher Education Funding Council for England. The SDSS
Web site is http://www.sdss.org. The SDSS is managed by the Astrophysical Research Consortium (ARC) for the Participating
Institutions. The participating institutions are the American
Museum of Natural History, the Astrophysical Institute
Potsdam, the University of Basel, the University of Cambridge,
Case Western Reserve University, the University of Chicago,
Drexel University, Fermilab, the Institute for Advanced Study,
the Japan Participation Group, The Johns Hopkins University,
the Joint Institute for Nuclear Astrophysics, the Kavli Institute
for Particle Astrophysics and Cosmology, the Korean Scientist
Group, the Chinese Academy of Sciences (LAMOST),
Los Alamos National Laboratory, the Max Planck Institute for
Astronomy (MPIA), the Max Planck Institute for Astrophysics
(MPA), NewMexico State University,Ohio State University, the
University of Pittsburgh, the University of Portsmouth, Princeton
University, the United States Naval Observatory, and the
University of Washington.

\begin{figure}
\epsscale{1.0}
\plotone{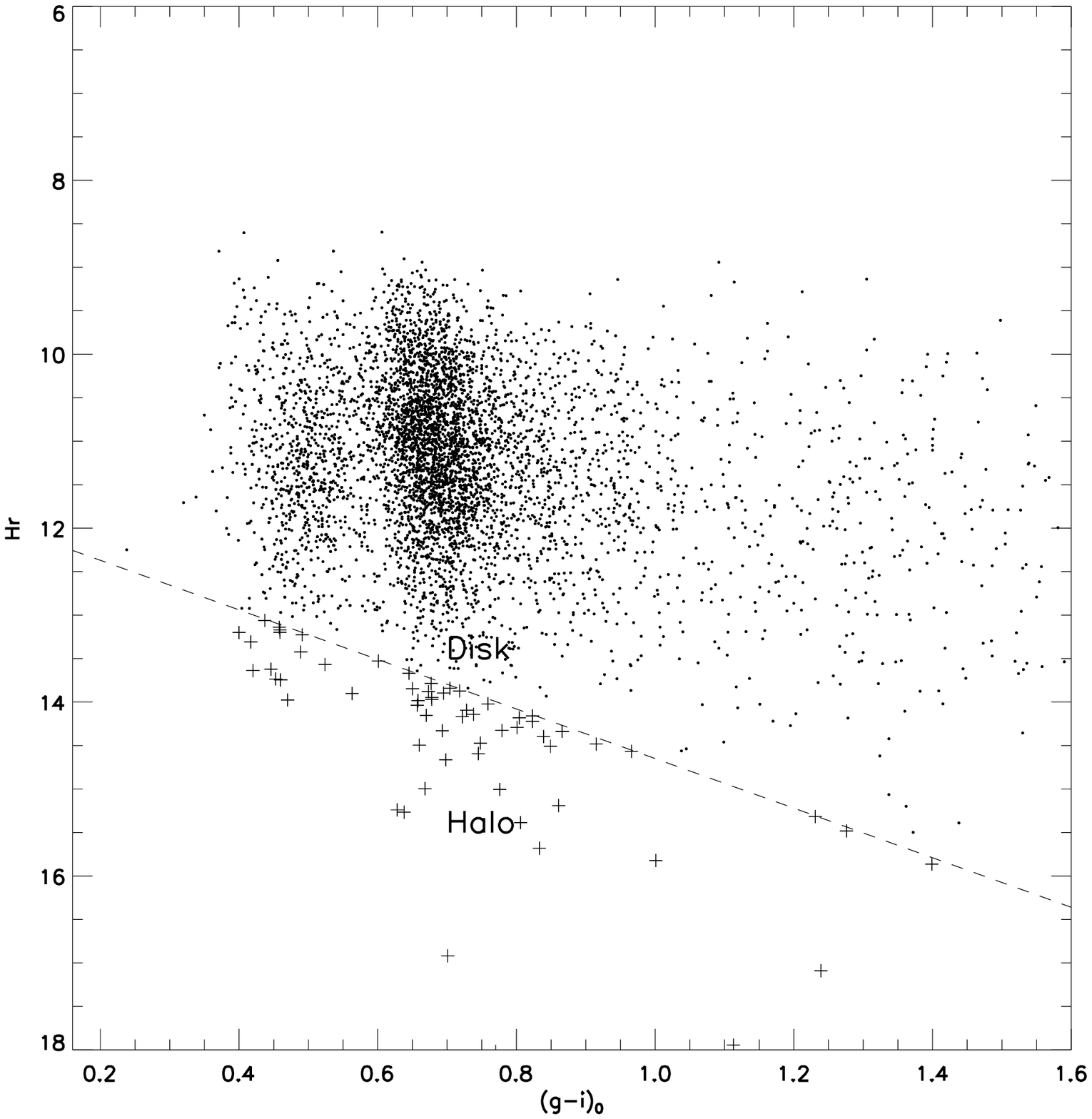}
\caption{Reduced proper motion (RPM) diagram of our sample. Hr $\equiv$ r+5*log$\mu$+5. The dashed line is the dividing line between disk and halo from Smith et al. (2009). Plus symbols indicate possible halo stars, which can be ignored because of the small amount.}
\end{figure}

\begin{figure}
\epsscale{1.0}
\plotone{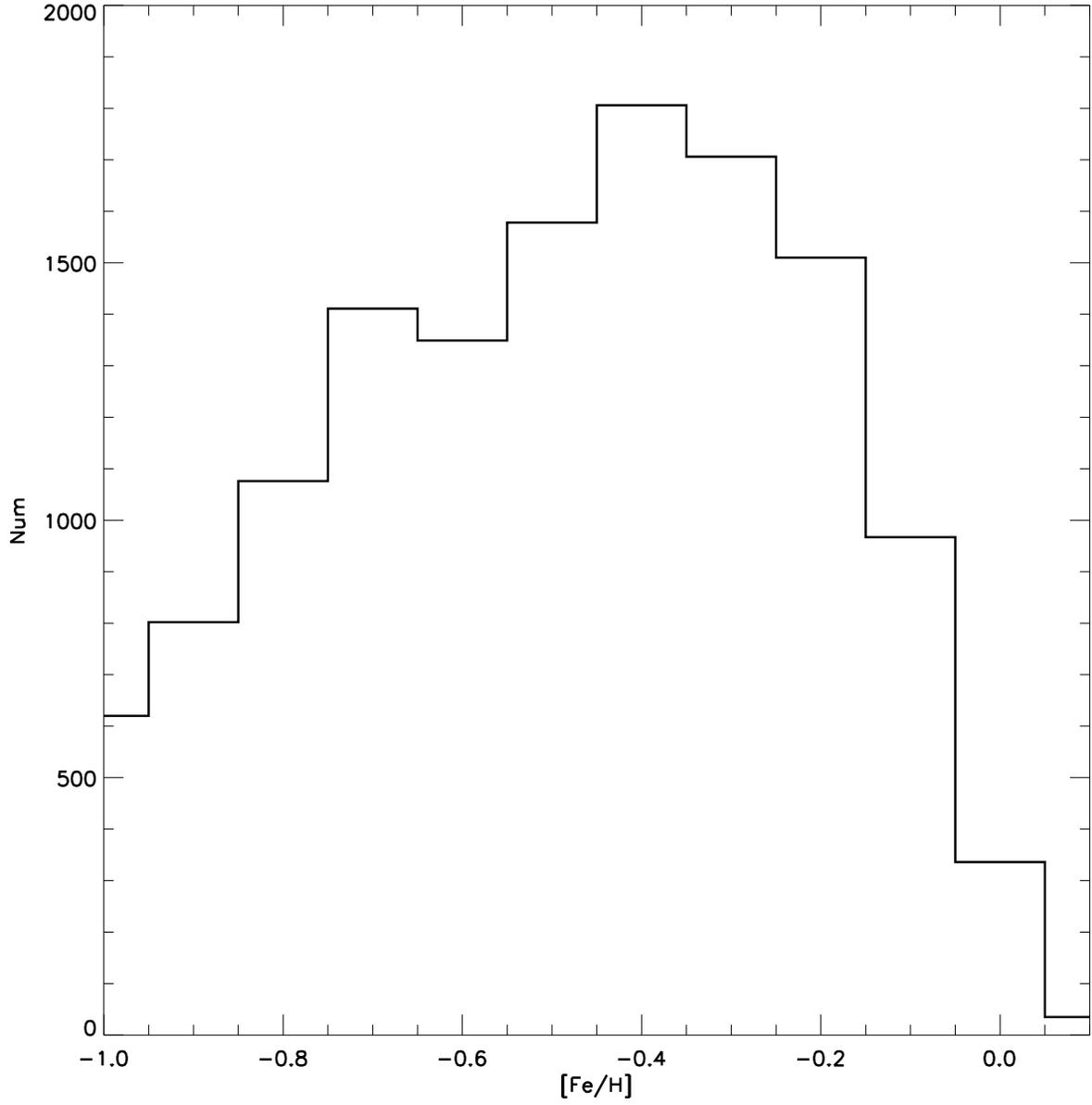}
\caption{Metallicity distribution of the total sample. The [Fe/H] peak locates at $\sim$ -0.4 dex. It should be noted that the metallicity distribution of our sample is a little different from that of other literature. }
\end{figure}


\clearpage
\begin{figure}
\epsscale{1.0}
\plotone{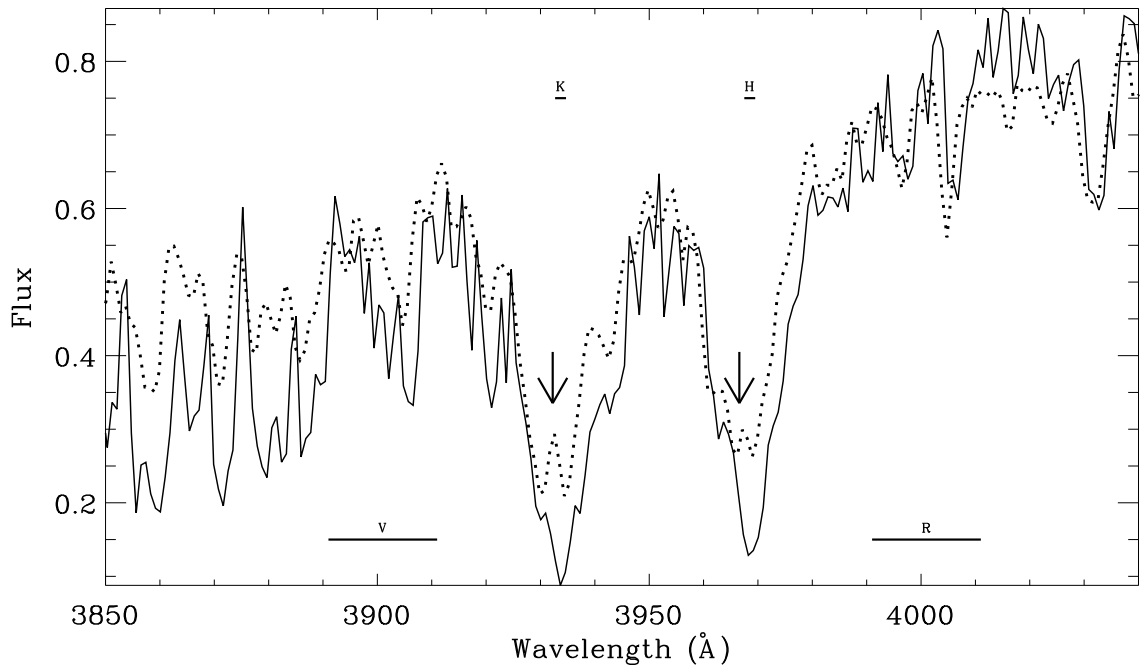}
\caption{Spectra of two stars. Solid line represents an inactive star while dotted line represents an active star. The emission reversals at the center of CaII H$\&$K are indicated by arrows. The width of the V, R, H and K are indicated by the horizontal lines under each index letter, respectively. }
\end{figure}

\begin{figure}
\epsscale{1.0}
\plotone{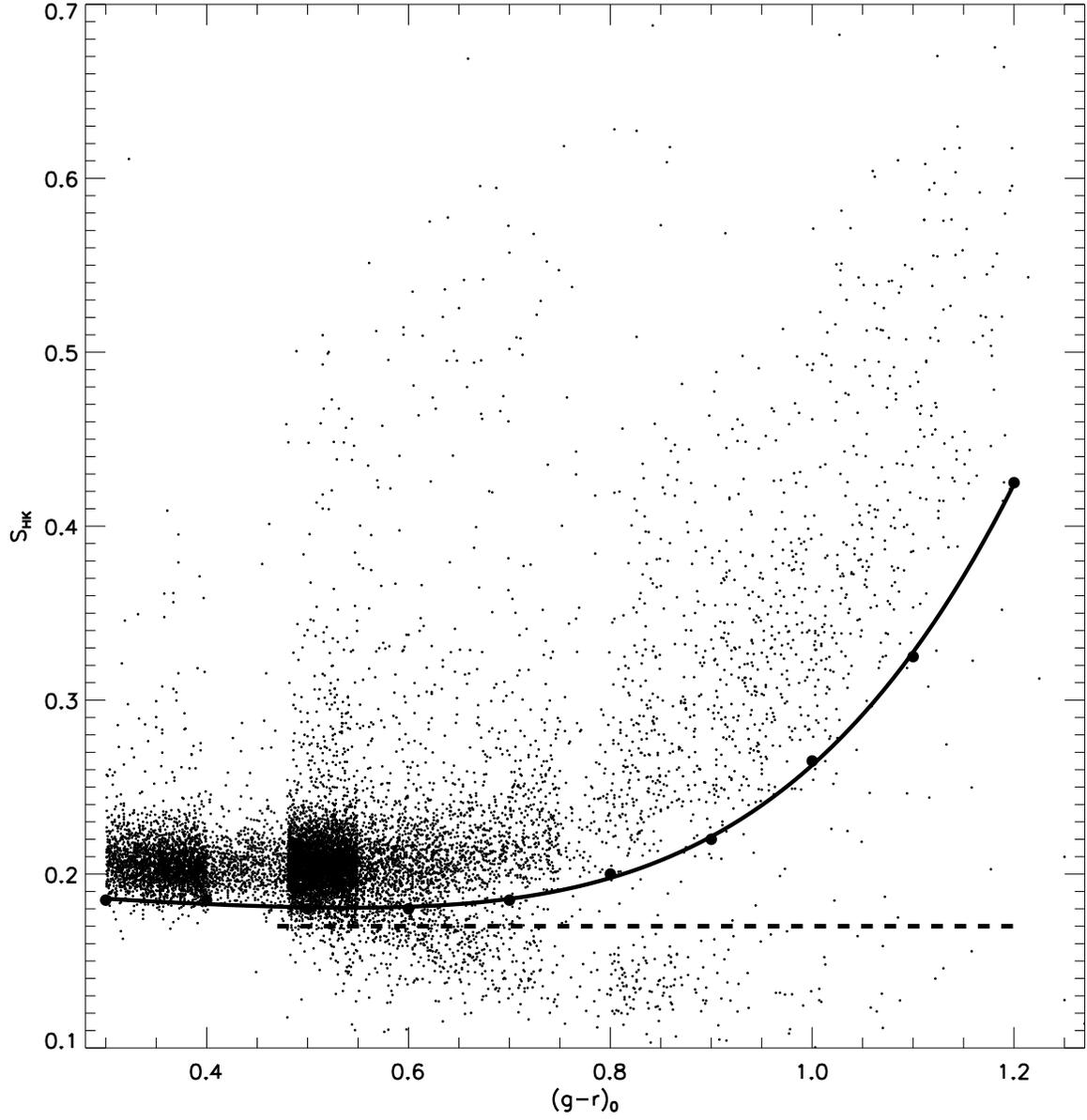}
\caption{$S$$\rm_{HK}$ vs. $(g-r)_{0}$. The black points represent field stars in our sample. The dashed line is the boundary line for very inactive stars, as defined in text, which tend to be subgiant and giant stars. The solid line is `zero' emission line for MS stars, as explained in the text. }
\end{figure}

\begin{figure}
\epsscale{1.0}
\plotone{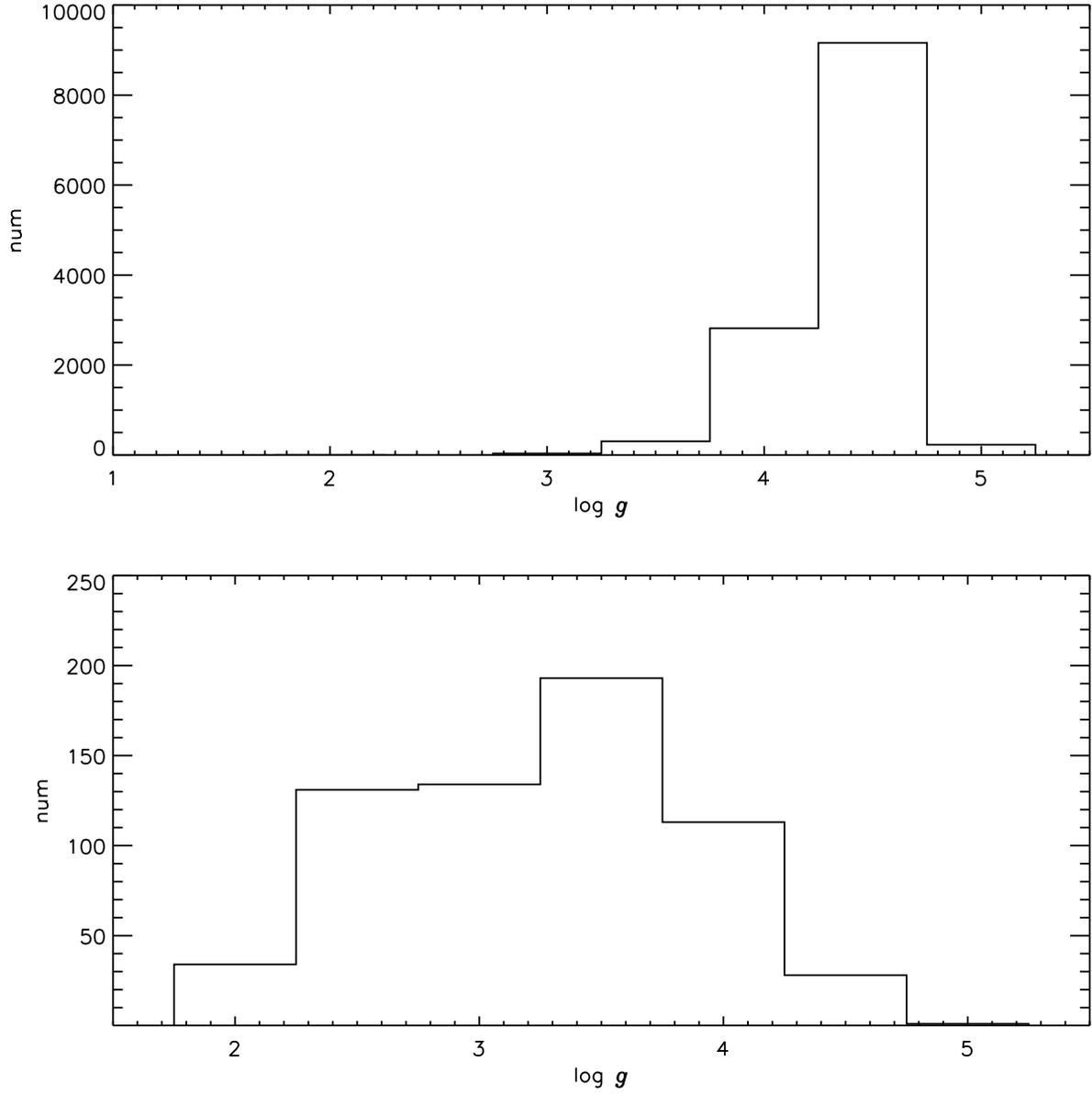}
\caption{Top: the log $g$ distribution for the stars with $S\rm_{HK}$ $>$ 0.17; Bottom: the log $g$ distribution for the stars with $S\rm_{HK}$ $<$ 0.17. }
\end{figure}


\begin{figure}
\epsscale{1.0}
\plotone{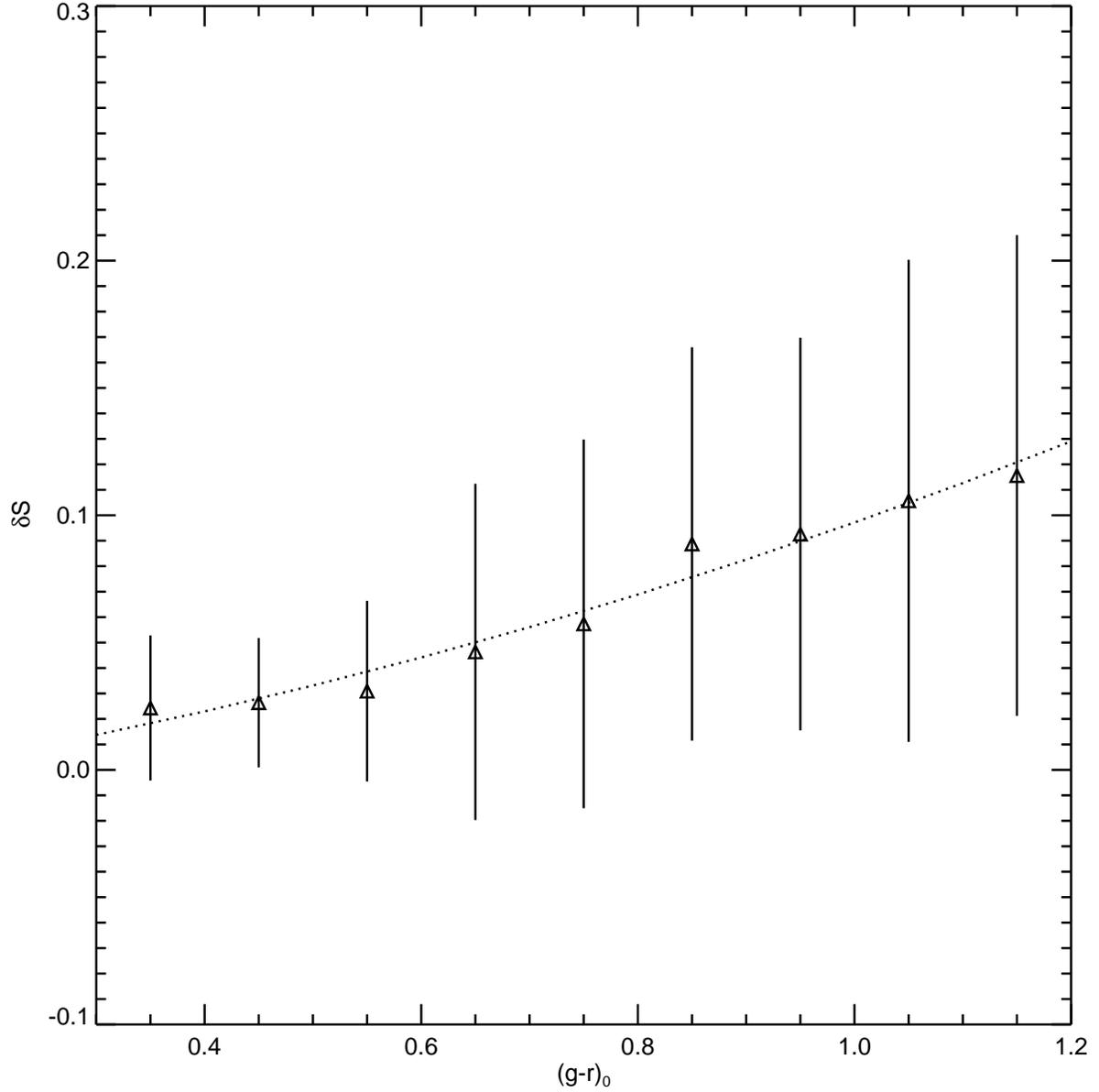}
\caption{$< \delta >$S vs. $(g-r)_{0}$ in color bins for dwarf stars. The triangles represent the mean $\delta$S value $< \delta >$S in each color bin. Error bars are the standard deviations in each color bin. Dotted line is the least squares fit of the average $\delta$S. Clearly both the range and mean CA is a strong function of color among MS stars.}
\end{figure}

\begin{figure}
\epsscale{1.0}
\plotone{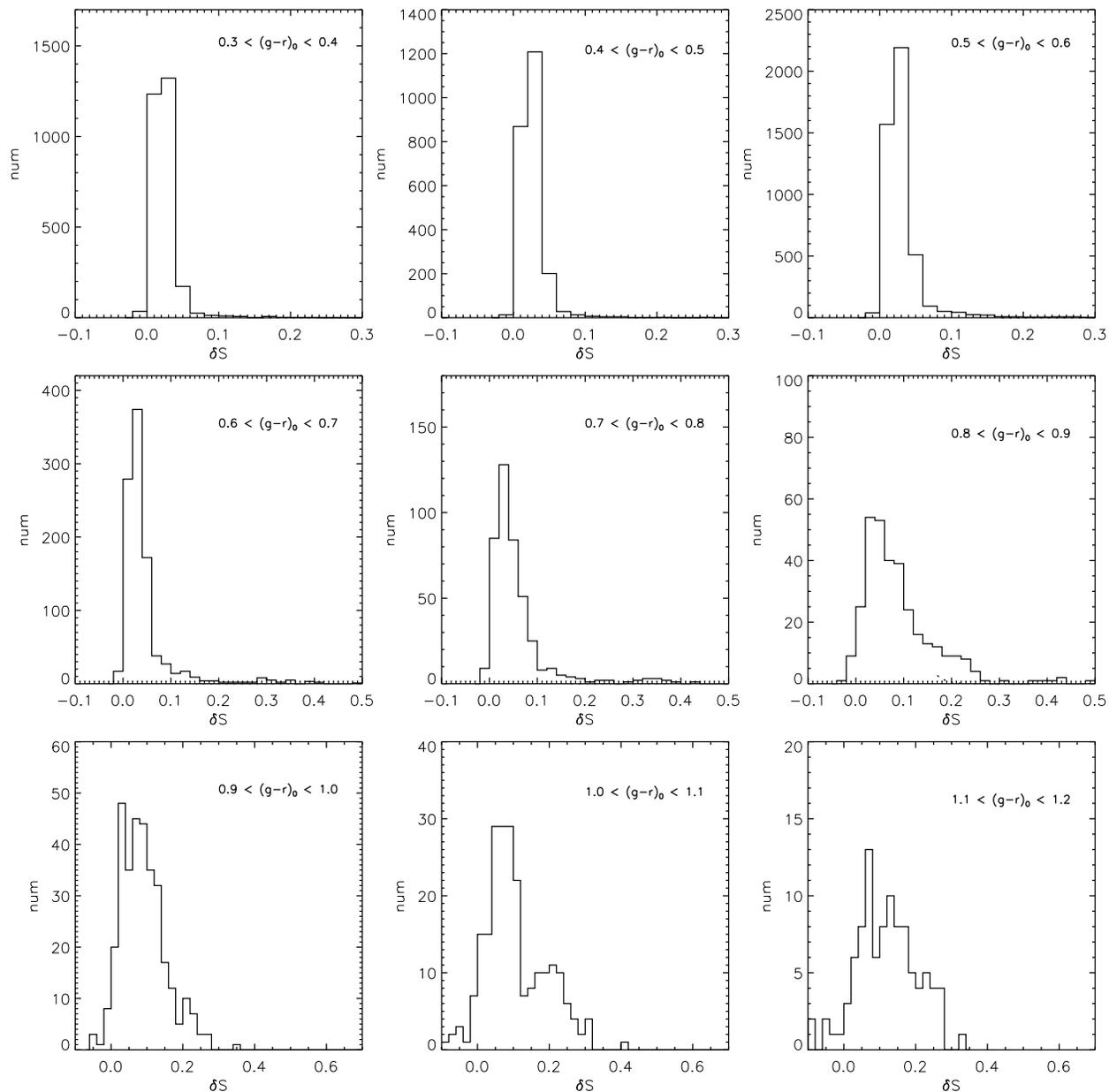}
\caption{$\delta$S distribution in different color bins. Within 0.3 $<$ $(g-r)_{0}$ $<$ 1.0 the distributions show only one peak.   Within 1.0 $<$ $(g-r)_{0}$ $<$ 1.2 (last two panels) the distributions are bimodal. }
\end{figure}

\begin{figure}
\epsscale{1.0}
\plotone{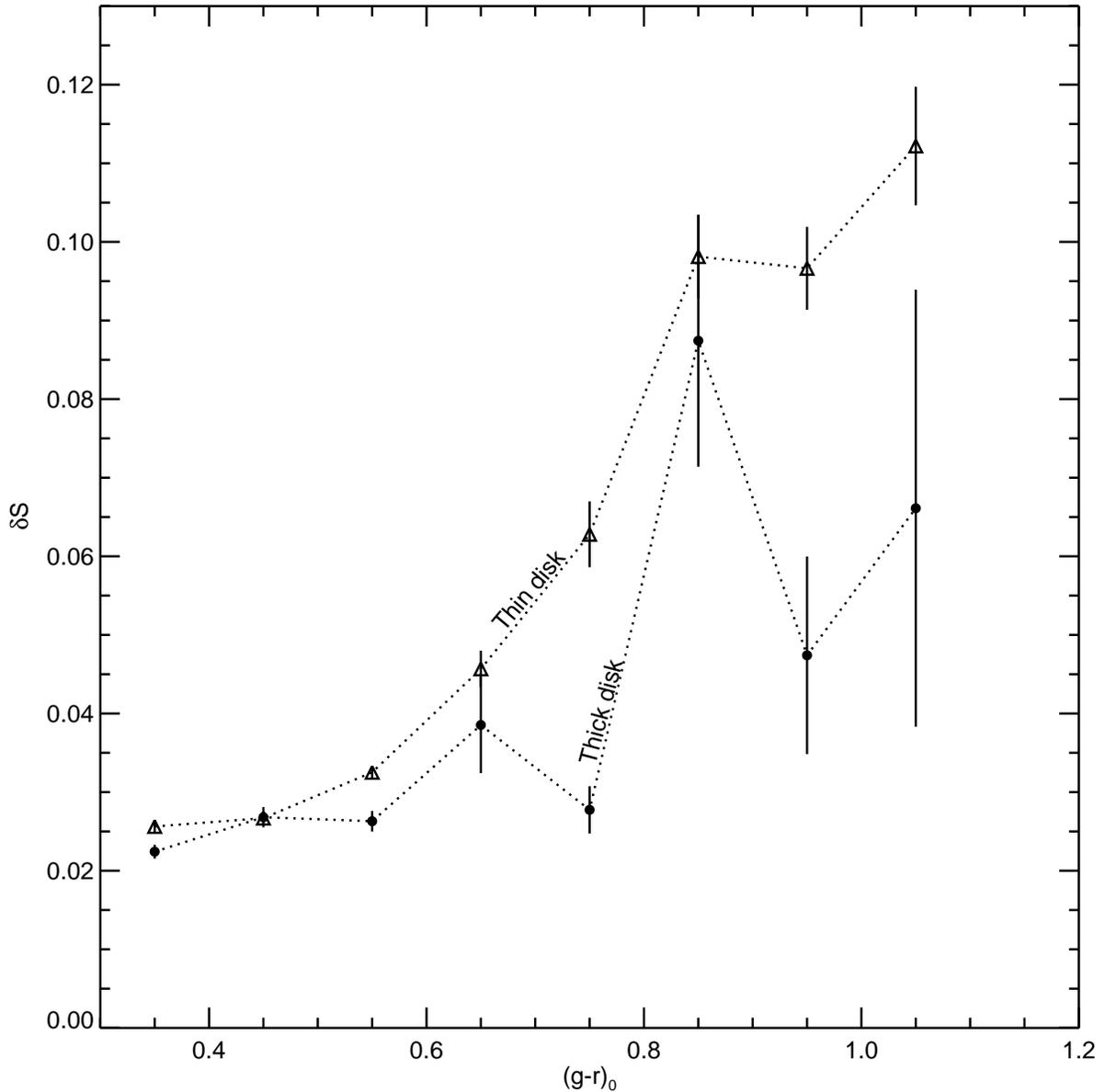}
\caption{$\delta$S vs. $(g-r)_{0}$ in color bins. The triangles represent the average $<$$\delta$S$>$ value for thin disk dwarfs in each color bin and the error bars are the mean standard deviations ($\sigma$m). The filled circles  represent mean $<$$\delta$S$>$ for thick disk dwarfs in each color bin and the error bars are the mean standard deviations. Clearly thin disk stars on average are more active than thick disk stars regardless of color. However, redder stars tend to be more active regardless of population.}
\end{figure}

%
%


\begin{figure}
\epsscale{1.0}
\plotone{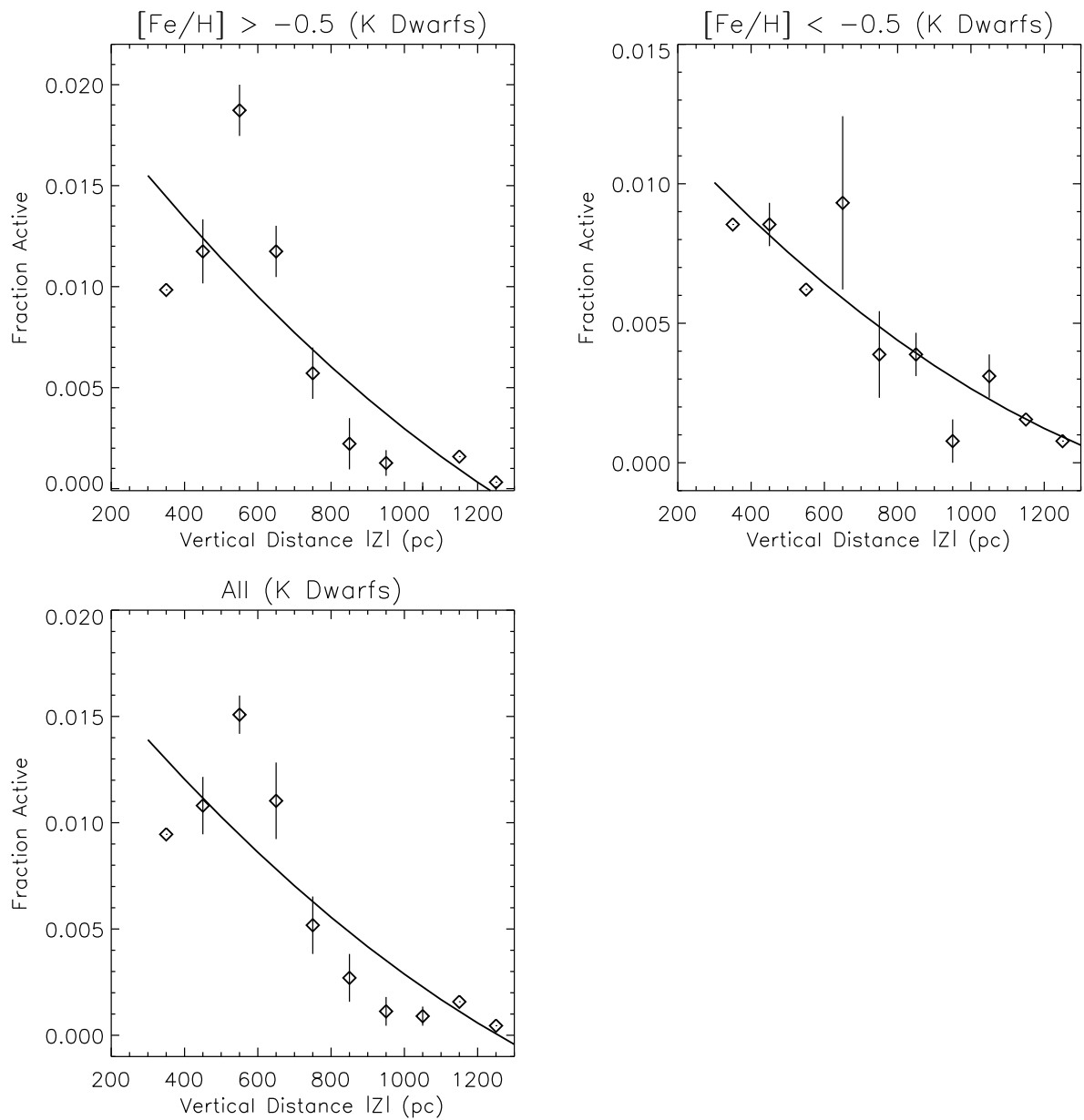}
\caption{The fraction of active stars vs. vertical distance among K dwarfs. The diamonds represent the fraction of active stars in each distance bin. Error bars are indicated by vertical lines.  The top left panel plots high metallicity stars with [Fe/H] $>$-0.5; The top right panel plots low metallicity stars with [Fe/H] $<$ -0.5; The bottom panel is for all the K dwarfs. The solid lines are the least square fits of the fractions.}
\end{figure}


%
%

\begin{figure}
\epsscale{1.0}
\plotone{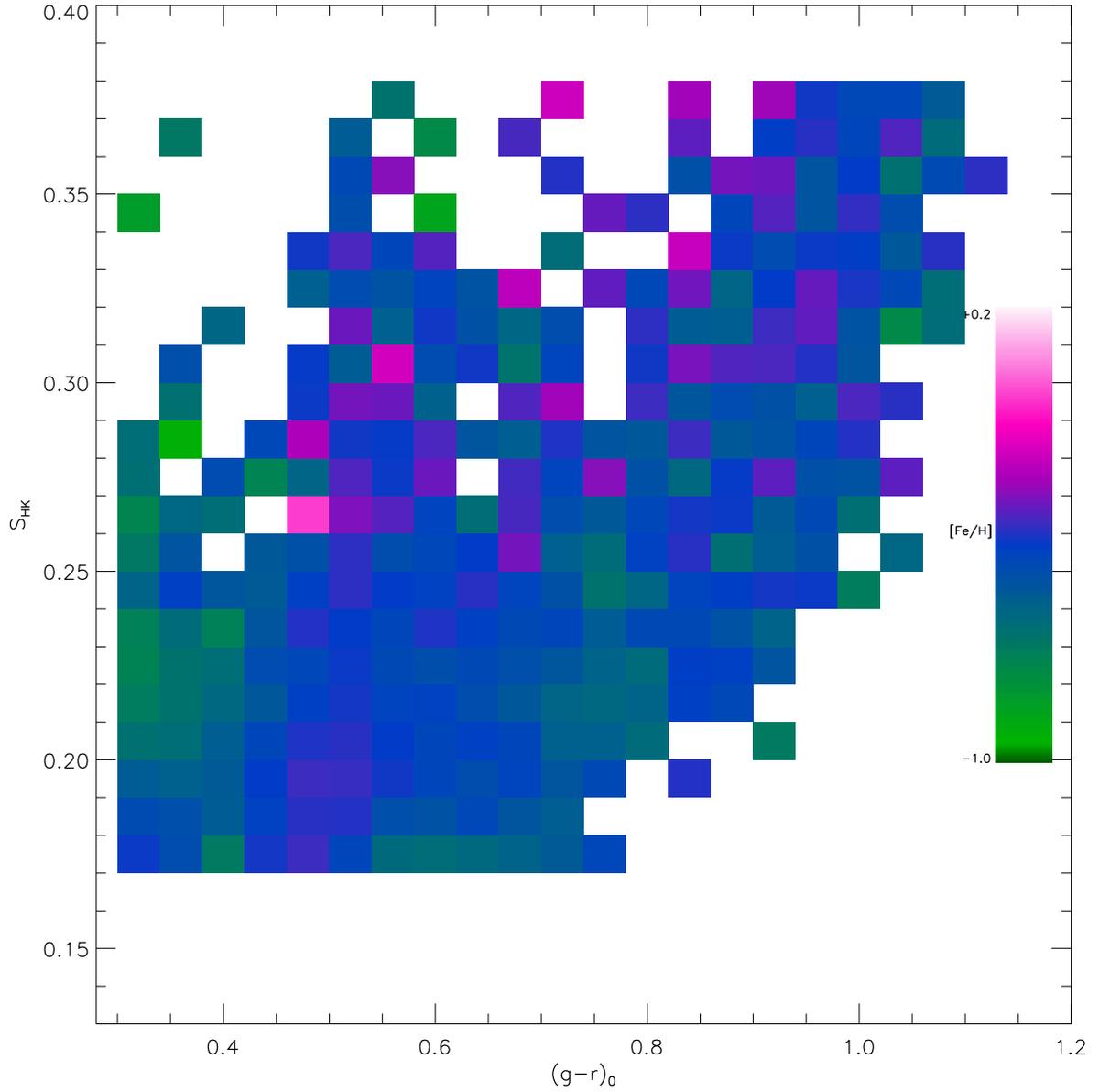}
\caption{$S$$\rm_{HK}$ vs. $(g-r)_{0}$. The bin size of each rectangle is 0.04 in $(g-r)_{0}$ and 0.01 in $S\rm_{HK}$. The color  represents the average metallicity in the rectangle.  The color bar indicates the relation between the color and the [Fe/H]. }
\end{figure}

\begin{figure}
\epsscale{1.0}
\plotone{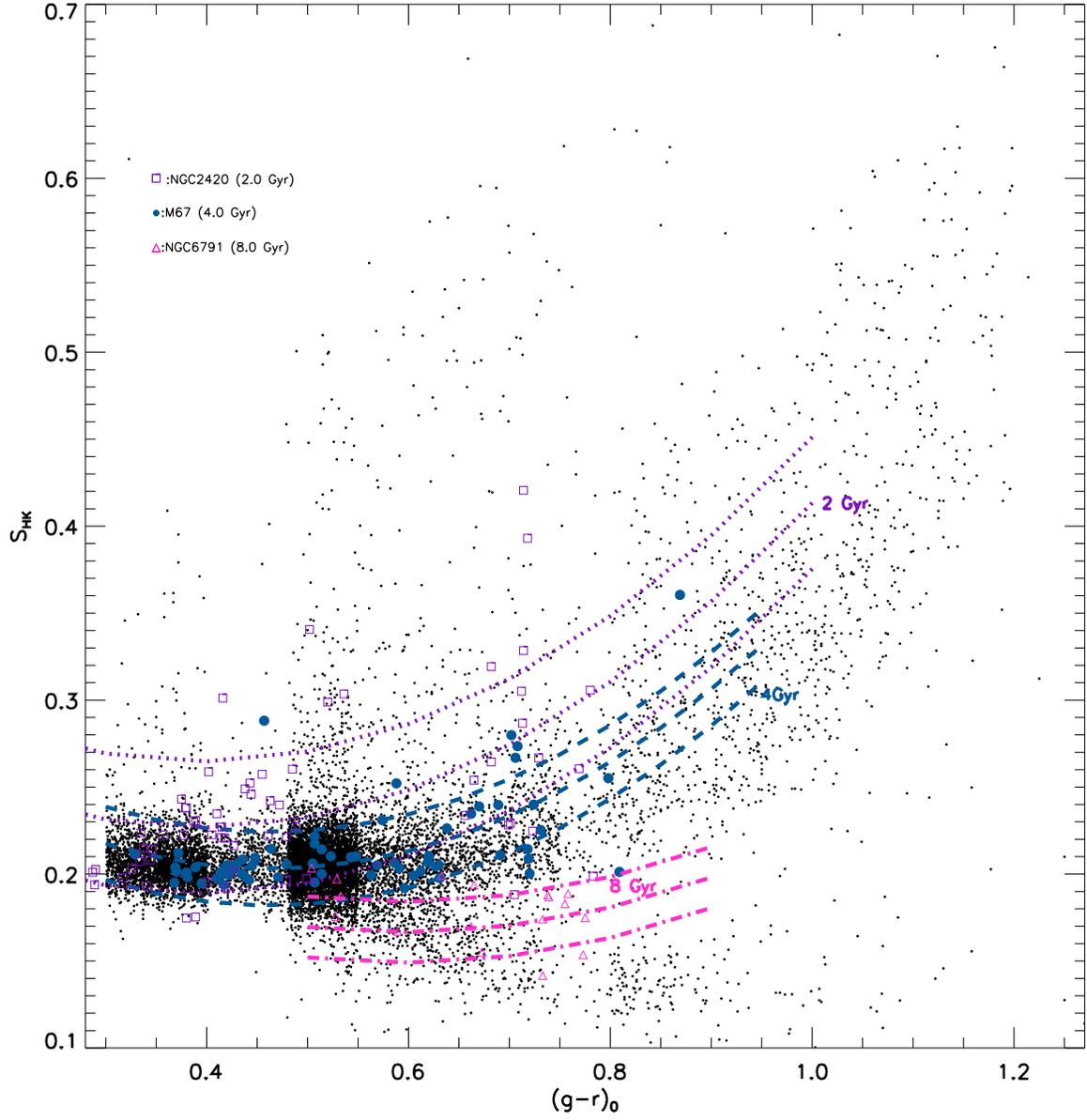}
\caption{$S$$\rm_{HK}$ vs. $(g-r)_{0}$. The black points represent field stars in our sample. The open squares, blue filled circles and open triangles represent the member stars of open cluster NGC2420, M67 and NGC6791, respectively. The dotted line, the dashed line and the dash dot line are the least square fits and $\pm$1 $\sigma$ of those three cluster.  }
\end{figure}




\clearpage

\clearpage




\end{document}